# Scientific Communication and Cognitive Codification:

# Social Systems Theory and the Sociology of Scientific Knowledge

Loet Leydesdorff

Amsterdam School of Communications Research (ASCoR), University of Amsterdam,

Kloveniersburgwal 48, 1012 CX  Amsterdam, The Netherlands

loet@leydesdorff.net ; http://www.leydesdorff.net

**Abstract**

The intellectual organization of the sciences cannot be appreciated sufficiently unless the cognitive dimension is considered as an independent source of variance. Cognitive structures interact and co-construct the organization of scholars and discourses into research programs, specialties, and disciplines. In the sociology of scientific knowledge and the sociology of translation, these heterogeneous sources of variance have been homogenized *a priori* in the concepts of *practices* and *actor-networks*. Practices and actor-networks, however, can be explained in terms of the self-organization of the cognitive code in scientific communication. The code selects knowledge claims by organizing them operationally in the various discourses; the claims can thus be stabilized and potentially globalized. Both the selecting codes and the variation in the knowledge claims remain constructed, but the different sub-dynamics can be expected to operate asymmetrically and to update with other frequencies.

**Practices and the Pansemiosis of Actor-Networks**

In his seminal study of the Sociology of Scientific Knowledge (SSK), David Bloor noted that 'knowledge for the sociologist is whatever men take to be knowledge' (1976: 2). Consequently, this 'strong program' in the sociology of science introduced a principle of symmetry into explanation: a sociological explanation in terms of human beliefs should be able to explain both true and false knowledge. From this perspective, scientific knowledge can no longer to be defined as 'true' belief, and therefore different from other knowledge (Barnes, 1974). In the case of mathematics, Bloor (1982) argued that even rules of logical inference derive their truth from social negotiation and human belief.

The strong program extended on Thomas Kuhn's (1962) notion of a paradigm as a language game (Winch, 1958; Barnes, 1969): in *practices* the cognitive is always social, and *vice versa*. The dimensions of the cognitive and the social are integrated and cannot be distinguished; socio-cognitive (inter-)actions shape the social *and* the cognitive at the same time (Collins, 1983). Therefore, analysis should not be pursued in terms of dimensions like 'cognitive' versus 'social' or 'internal' versus 'external' (Callon *et al.,* 1983). SSK has not accepted any *ex ante* disciplinary division of labor among the history, philosophy, or sociology of science in terms of their subject matter.

In comparison with older traditions in the sociology of science (Merton, 1942; 1973), this focus led to descriptions of the world of science that were empirically richer than those provided by more traditional approaches in sociology and philosophy. For example, it was no longer possible to describe a specialty only in terms of the organizational variables of a scientific community (Crane 1969 and 1972; Whitley 1984). Nor could a specialty be operationalized in purely epistemological terms as a set of theoretical questions linked to relations among observations, arguments, and inferences (Hesse 1980); nor could it be adequately described as a body of literature or a communication structure (Price, 1961). As with all major concepts in science studies, it was henceforth necessary to develop the definition of 'specialty' from the perspectives of social structure, cognitive structure, and scientific literature. The potential tensions among these different evaluations were 'heterogeneously engineered' in *practices* which operate as 'mangles' (Pickering, 1995); the analyst remains no option other than 'following the actors' as an ethnographer (Latour, 1987).

Scholars working in the tradition of Actor-Network Theory (ANT) radicalized this position by including non-human elements in the description (Callon *et al.*, 1986; Callon and Latour, 1981). For example, in a study of the introduction of scientific principles of breeding into fishery, Callon (1985) argued that the actor-network consists of the oceanologists who try to transform fishing into 'aquaculture,' the science of oceanology which imposes a problem-formulation, the fishermen who defend their interests, and the scallops who breed and enter the networks. When all these elements interact, the system can be 'translated' because an 'obligatory passage point' is generated. Note that in this 'sociology of translation' the cognitive or natural constraints on the situation are not analyzed *as if* they acted upon the situation—that is, as an heuristic device; every unit should instead be analyzed in substantively similar terms, that is, as semiotic 'actants' in a network. The actor-network is constructed as a next-order unit of performance (containing 'relational strength'), into which the heterogeneous dimensions are homogenized.

In other words, the substantive heterogeneity in the *explanandum* is not addressed in terms of analytically different dimensions, but in terms of an assumed coincidence and congruity between *explanandum* and *explanans* within the subject matter. The actor-networks cannot be explained other than by describing them as empirical phenomena. Since one knows *a priori* that the relations in the actor-network are mutual and symmetrical, nothing can ultimately be explained, and the sole purpose of the analysis is to tell a story (Latour 1987; Collins and Yearley 1992). Consequently, the actor-network is not only an *empirical* category; it is also an answer to the *methodological* problem of analyzing 'heterogeneity.'

**The Status of Cognitive Structure**

Three serious attempts have been made to break this vicious circle of equating the *explanans* with the *explanandum* in the sociology of scientific knowledge and actor-network theory. First, Peter Slezak (1989) opened a symposium in *Social Studies of Science* with the claim that the reproduction of scientific discoveries by the computer, that is, without a given context (e.g., Langley *et al*., 1987), has shown that the cognitive puzzles of the sciences may be conditioned by social relations, but are not necessarily or significantly determined by them. In his opinion, social factors would not be sufficiently fine-grained to accommodate the details of cognitive distinctions in science (Schmaus *et al*., 1992).

The standard reply to this critique invokes the Quine-Duhem thesis, which states that theories are underdetermined by empirical evidence, and that there is room for social factors and social explanations within this cognitive uncertainty (Pinch, 1985). However, this reasoning implies a *non sequitur* (Slezak, 1989: 588f.). The underdetermination concerns the formal relation of any theory to its direct observational basis and is still neutral on the issue of social determination (Dorling, 1979).

Thus, the question about the relation between the cognitive and the social dimension can be turned into an empirical one: which variation is to be explained, and which *explanans* can provide the explanation? Note that this relation itself may vary in different stages of the historical development of a paradigm. However, explaining a large (cognitive) variance in terms of a smaller (social) one may lead to trivial results in an increasing number of case descriptions:

> Simply multiplying the output of case studies and citing the growing 'weight' of this evidence, as is often done, fails to improve the situation. This stock response entirely misses the point. Thus, Shapin complains that 'The continued assertion that scientific knowledge is autonomous and transcendent is now more often founded on ignorance of this new literature than on considered criticism thereof.' (Slezak, 1989: 585).

The methodological perspective—of variances which are to be explained in terms of other variances—provides us with a second line of critique of SSK and ANT. When different dimensions interact and co-vary in the observable events, they determine each other in their mutual interactions, but condition each other *asymmetrically* in the remaining variances. Consequently, uncertainty in the cognitive dimension is composed of an uncertainty that is *determined* by the social variation and a complement that is only *conditioned* by the latter. If one focuses only on the (symmetrical) determination in the interaction, the remaing variances are no longer estimated.

For example, when an author publishes a text, the cognitive and the social interact in this event. The publication may improve the author's social standing in the department, but its impact at the field level of the relevant scientific community is evaluated in terms very different from the institutional ones. Scientific communities provide reputations and prizes, and not promotions or tenure like university departments. Thirdly, it matters in both other dimensions whether the text was a research article, a review, or a book-length monograph.

In my study *The Challenge of Scientometrics*, I argued that three dimensions can be distinguished as crucial to the study of the sciences: authors, texts, and cognitions (Leydesdorff, 1995). These three units of analysis cannot be reduced to one another: the variation in each of the dimensions can be aggregated using grouping rules in the other(s). Thus, the following scheme can be developed:

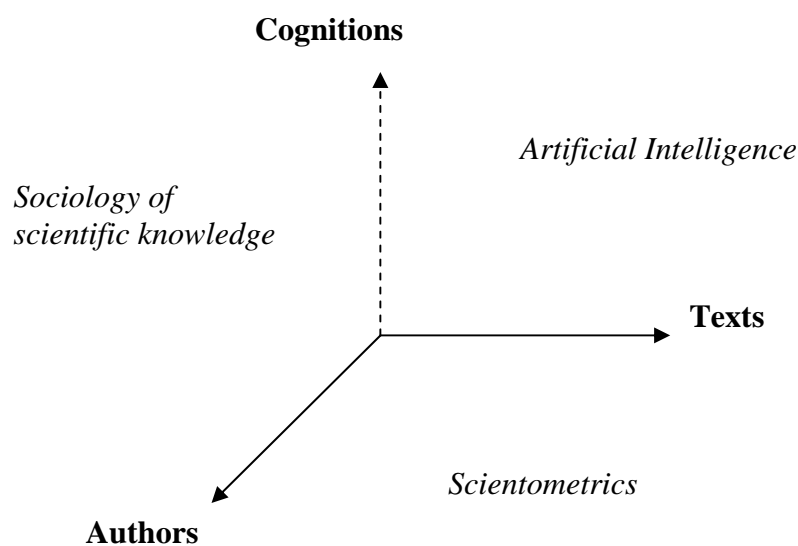

**Figure 1**: An analytical scheme for studying the sciences

In this scheme, the line indicating 'cognitions' is dotted because, unlike texts and authors, cognitions cannot be found 'out there' without taking a *reflexive* turn. Still, the cognitive dimension provides important grouping rules for organizing texts and authors intellectually into research programs, specialties, and disciplines.

Note that the assumption of *analytical* independence of the cognitive dimension does not imply a return to older traditions in the philosophy or sociology of science. Traditionally, the two dimensions have been conceptualized as two separate domains, to be pictured in a spatial metaphor as parallel planes (the context of discovery and the context of justification) and to be studied by distinctive scholarly traditions. However, by considering the analytical dimensions as orthogonal, the scheme unfolds a multi-dimensional construct in which it is possible to specify different meanings—based on different grouping rules—for the very concepts of science studies. This model allows additionally for 'interaction' (at each moment) and for dynamic 'feedbacks' and 'mutual shaping' among the dimensions over time.

In order to address the cognitive dimension, the issue of reflexivity is pertinent. How can cognitions emerge both within and among agents, and be sedimented also into texts? In a third stream of critique, Mulkay *et al.* (1983) argued that the initial assumption of 'symmetry' in the strong program, that is, symmetry in explaining true and erroneous beliefs, is untenable if one studies the sciences empirically as forms of discourse. Scientists apply at least two repertoires: one of them is contingent, and the other these authors call 'empiricist.' Significantly, beliefs taken to be correct are expressed more in the empiricist repertoire, for example, in formal scientific literature, while scientists tend to use contingent repertoires in order to account for allegedly incorrect beliefs:

> Criteria are presented as constituting a clear-cut, impersonal, unavoidable constraint on the choice of correct theories; whilst the same criteria are much more likely to be depicted as socially contingent and malleable when they are cited in connection with incorrect theories. (*Ibid.*: 198.)

Why do scientists *asymmetrically* attribute certain types of beliefs and the corresponding verbal behavior to certain arguments and not to others? Why do they act as if they believed in science? Once one reintroduces 'asymmetry' into the explanation of beliefs deemed correct versus those considered erroneous, the important next question is whether the fact that a belief is held to be correct may serve also as a useful indicator for the correctness of this belief. However, the question of the warrant for a given belief has been the central question in the philosophy of science, and therefore one either has ultimately to accept the *not exclusively* social (but epistemological) character of the issue, or one can radically adopt a (hyper-)reflexive position in which in the end there is nothing but belief, and therefore no such thing as 'objective science' is possible (Woolgar 1988).

Mulkay *et al.* (1983) noted that scientists involved in controversies tend to raise the proportion of contingent elements in their discourse, while 'the discourse forthcoming from less lively fields has proved to be comparatively intractable to sociological investigation' (p. 198). These authors, however, did not infer from this variation in the phenomena the possibility of another (e.g., cognitive) factor which might explain the difference. Is the communication coded differently depending on whether the problem definitions are 'hot' or 'cold'? (Callon, 1998). How does language enable us sometimes to construct 'objective' (or intersubjective) meaning for 'true statements' and at other times to deconstruct this objectivity in terms of contingent circumstances? Let us take a more precise look at the social construction of scientific knowledge in language.

**Codification in communication**

When a scientific paper is presented at a conference, its content has only the status of a knowledge claim. However, when the paper is subsequently reviewed by peers and published in the literature, this knowledge claim is validated, and thus the epistemological status of the content is changed by the operation of the relevant social system (Myers, 1985). Peer review is, among other things, expected to check the paper under study for its quality, and while doing so it inscribes an expectation of quality into the paper as a construct.

In his study *Die Wissenschaft der Gesellschaft*, Niklas Luhmann (1990a) argued in favour of shifting attention away from the social process of construction by agency and towards this change in the status of the constructed. What has been added to the article during this social process? Building on Parsons's (1963a, 1963b) concept of symbolically generalized media of communication with specific codes, on the one hand, and on Husserl's (1929) notion of a horizon of meanings, on the other, Luhmann suggested that the coding of the communication implies a domain-specific selection. This expectation of a selection allows the participants in the communication to handle more complexity by focussing on the content of a claim.

In other words, the paper goes through a process of selection whereby it is invested with symbolic value. The cognitive code of the communication is characterized by Luhmann as a selection on whether the content of the paper is also true. Although these standards themselves are also constructed historically, they operate as selection mechanisms at another level than the knowledge claims that are selected.

The construction and differentiation of the selecting code as a basis for a scientific culture has taken centuries. For example, the scientific journal is an invention of the 17th century (Price, 1961), while the modern citation was invented only at the end of the 19th century (Leydesdorff and Wouters, 1999). Cognitive criteria operate globally as horizons of possible meanings which are reproduced locally by the specification of the code of communication under historical circumstances, that is, by accepting or rejecting the knowledge claim in each paper under review (Fujigaki, 1998).

This notion of 'horizons of meaning' was taken by Luhmann from Husserl's transcendental phenomenology (Paul, 2001). In his *Cartesian Meditations* of 1929, Husserl specified intersubjectivity and communication as a monade different from the psychological one. Only by placing the psyche between brackets—in a so-called epoché (εποχη)—is one able to uncover in oneself (through meditation and *not* through discourse)[1] the double contingency in one's relation with fellow human beings: *Ego* not only encounters the other as another subjectivity (which is given), but expects *Alter* to entertain expectations similar in nature but potentially different from those one is able to perceive in one's own *Ego*.

---

[1] The *noesis* precedes the discursive construction.

The exchange of expectations in interactions generates meaning at a supra-individual level. Meaning can further be codified. Knowledge, for example, can be considered as a meaning that makes a difference. In the case of discursive knowledge, this difference is defined with reference to a code in the communication which itself remains uncertain and contingent. For example, the code can also be redefined by the discourse in the case of a crisis in the communication. Thus, the network of communication develops an eigen-dynamics which is partly (that is, reflexively) accessible and partly latent for the communicators who carry the communication (Lazarsfeld & Henry, 1968; Von Foerster, 1982).

Luhmann took from Parsons (1963a and 1963b) the idea that the codes of the communication can be generalized symbolically and differentiated functionally. For example, transactions on the market place are guided by a code that is very different from the code governing scientific communications. However, Parsons did not combine these two elements of his theory of social systems. In Parsons's theory, actions are differentiated with reference to functions for the social system and not as communications with reference to codes. Furthermore, Luhmann (e.g., 1986) added to social-systems theory the theory of autopoiesis or self-organization, which he took from Maturana and Varela (1980). Using the concept of autopoiesis, differentiation can be considered as an endogenous result of the codification of communication. The communication self-organizes in terms of its meaning, and this relation is recursive. When meanings can again be communicated, differences among them can also be codified.

**Beliefs and Rationalized Expectations**

How does this thesis of functional differentiation among the codes of communication lead to a sociology of science that is different from the sociology of scientific knowledge discussed above? The strong program sought to explain all cognitive variation in terms of its socio-cognitive construction. This programmatic assumption was introduced into science studies by Bloor (1976: 40ff.) with reference to Durkheim's (1912) analysis of the forms of religious life. However, beliefs remain attributes of agents or groups of agents. The sciences have been socially constructed as *discursive systems of rationalized expectations* not only in terms of their daily operations as practices, but in a much more profound sense, that is, in their relation to society at large, and notably with reference and in opposition to religious systems.

While a belief can be attributed to a community of people (e.g., a church), rationalized expectations are attributes of a discourse. In contrast to a system of expectations, a belief system are normatively integrated and hierarchically organized with reference to a codified meaning (e.g., religious Truth). Thus, the units of analysis are different: beliefs belong to the nodes of a network, whereas scientific communications are based on selections among the links. Perhaps the hypothesis that the sciences also function as belief systems can be empirically fruitful (e.g., in controversy studies), but case studies can never disprove that the sciences operate in important respects very different from belief systems. The analogy misses the point that as a scientist, one is free to theorize in modern societies, that is, that scientific discourses can also be considered as functionally differentiated subsystems of communication.

For example, the disbelief in a scientific 'truth' no longer necessarily creates a schism between two religious communities, as in the Middle Ages; nowadays it often raises only a variety of further research questions. Thus, the mechanism for the communication of uncertainty is different. Modern sciences are not hierarchically organized belief systems, but at least to a certain degree, and at the same time, they are juxtaposed discursive constructions. However uncertain and variable the coding may be, the methodological yardsticks for controlling the truth of scientific statements are functionally different from normative integration into individual or collective belief systems.

The differentiated media of communication can be considered as delocalized network functions; in a secularized society, normative integration can even be considered as yet another recursive network function, namely the specific one which has to be carried by localized agency (Maturana, 1978). The different functions tend to stand orthogonally in the model (Simon 1969). One can expect hierarchical stratification to be increasingly replaced with functional differentiation by the communication, since the spanning of a multi-dimensional framework (that is, in terms of the structural eigenvectors of the network instead of the relations of the agents) allows for the processing of greater complexity. Entirely new mechanisms of social coordination can also be induced.

Scientists have had a particular need for functional differentiation, since they need room for provisional interpretations or hypotheses that they may wish to change with hindsight. In the longer run, the sciences can allow for normative control only over the conditions of the communication (e.g., resource allocations), but not over the substantive and reflexive contents of these communications. Thus, the differentiation from normative integration has been a functional requirement for the further development of natural philosophy, that is, the new sciences. This crucial conflict was fought in Western Europe between the appearance of Galileo's *Dialogo* in 1632 and the publication of Newton's *Principia* in 1687. From that time onwards, functional differentiation has been further institutionalized in the social system (Leydesdorff 1994).

Why was the new philosophy of the 17$^{th}$ century able to drive this development? By reconstructing 'nature' in an experimental setting, an observation is transformed into an instantiation with reference to an expectation. Insofar as this reconstruction proves successful, that is, historically stabilizable, nature can then also be *replaced* with the new representation (Shapin & Shaffer, 1985). Consequently, the one paradigm can overwrite the other in new practices. In terms of the model, this replacement may be an evolutionary event or a gradual development. Once the old paradigm is replaced, it tends to lose also its theoretical meaning.

For example, it is difficult for us to understand why Huygens rejected Newton's concept of 'gravity' as completely 'absurd', while he was otherwise impressed by the *Principia*.[2] Equally, we no longer understand *why* it seemed important to Medieval physicians to let sick people bleed. Nowadays, we understand 'gravity' and 'blood pressure' as intuitively meaningful concepts. This translation mechanism in the reconstruction has driven cultural evolution in the course of the past few centuries. Scientific discourses provide the other subsystems of society with a reflexive window on the uncertainty inherent in representations. By explaining events as instances of ranges of possibilities, recombinations can be invented that may be better adapted to new contexts than the developments presumed to have occurred 'naturally.'

---

[2] Letter to Leibniz, 18 November 1690 (C. Huygens, *Oeuvres Complètes*, Vol. IX, p. 538).

The differentiation between normative expectations and cognitive ones, and the organization of other subsystems of social coordination in accordance with this differentation (that is, between normatively integrated belief systems and rationalized systems of scientific expectations) triggered the conflict of which the position of Galileo has become symbolic. The breaking of the hierarchical order of 'Rome' transformed the prevailing system from one based on stratification as the leading principle into one based mainly on functional differentiation among the codes of communication: the eternal Truth of the church provided by the Book and *ex cathedra* stands perpendicular to the provisional character of scientific knowledge claims.

**The Constructedness of the Epistemological Code**

In Luhmann's writings one finds a tendency to reify the functional differentiation of society and to interpret the codes of communication using a biological metaphor for the coding (Künzler, 1987; Leydesdorff, 2000, 2006). For example, Luhmann (1984) assumes that 'systems are given' (p. 30). At other places, however, he specified that the 'reality' under study remains unknown because the representing constructs stand tangentially to the represented systems (Luhmann, 1990b; cf. Casti, 1989). Thus, the systems under study are accessible as expectations entertained in the reflexive systems of communication. Scientific discourses generate standards endogenously which function as codes of communication, but these codes themselves also remain discursive constructs. They operate as methodological programs on the substantive programs, and this coevolution and mutual shaping can develop along a trajectory and thus stabilize otherwise volatile representations (Luhmann, 1990a).

Since communication systems contain uncertainty—because of their inherent distributedness among human beings—the codes of communication cannot be completely closed in their operation; they remain dependent on an uncertain operation. As subsystems of a social system, the codes disturb each other in interactions at the margins. Herbert Simon's (1969) model of complex systems as 'nearly decomposable' improves upon Luhmann's notion of 'operational closure.' While in biological systems operational closure can be structural, in social systems translations among codes remain possible.

For example, blood is not supposed to pass the blood-brain barrier in the body, since the two systems are operationally closed to each other. However, the social system should not be reified as if the hypothesized barriers existed physically: the different codes are expected to operate orthogonally within the model, but in historical time different codes can interact and thus be integrated locally into specific organizational formats. For example, publishing an article has meaning both socially and cognitively. These meanings, however, are quite different.

Luhmann's 'given' systems and their operational closure remain counter-factual, that is, defined only in the analytical model. But since this model is entertained by a discourse, it is also unavoidably a part of social reality. Can this distinction between 'social reality' and the model still be made if social reality itself—following

Husserl—is considered as an order of expectations? Without expecting expectations—as I specified above for the case of 'double contingency'—the social system cannot even be specified. Thus, the social system contains a *constructed* model of itself that enables us to process and change meaning at this level. While each psychological agent processes meaning naturally, one cannot change meaning at the level of the social system without having access to a horizon of meanings. Intersubjective meaning is no longer given 'naturally' nor does it follow a 'natural' trajectory (like the 'natural' preferences' of 'rational' actors); meaning emerges in social interaction from the analytical perspective of hindsight as a specific construction among other possible ones.

At the biological level, a signal can be distinguished from noise because of autopoiesis, which operates selectively, but at this level the observing unit which makes the distinction in its autopoiesis remains embedded as a node of the network. Language has provided us with an evolutionary tool to externalize observations as statements and thus to specify expectations. A statement provides information with meaning. The symbolic generalization of codes of communication within natural language, and their subsequent differentiation in terms of functions of communication, provides us with next-order communications which remain embedded in the substantive communications from which the codifications emerge.

While the systems are constructed bottom-up, control can be expected increasingly to invert the axes into top-down feedbacks by evolutionarily subsequent systems. However, communication systems should not be reified, because as operations they remain orders of expectations. The next-order niches within scientific discourses enable us to reduce the complexity with orders of magnitude. Cognitive variance proliferates much faster than social retention mechanisms can follow. New resources can thus be mobilized, for example, at the level of the motivation of the participating communicators.

Each functionally differentiated code of communication can be expected to operate in other (nearly-orthogonal) terms. While scientific communication has developed under the ethos of institutional disinterestedness (Merton, 1942), political communications and profit-seeking behaviour on the marketplace are driven by the explicit interests of the agents involved. However, these differentiated media permeate one another in actual events.

For example, an economy as a social subsystem of communications operates in terms of transactions on the marketplace. The markets function as clearing mechanisms which counteract imbalances between supply and demand. Innovations upset the equilibria (Schumpeter, 1939), and when the innovations are knowledge-based this upsetting can become structural (Nelson & Winter, 1982). The two selection mechanisms—that is, the one of the market at each moment of time, and the other along the time axis in a scientific discourse—stand in orthogonal relation, but since they operate in terms of communications and hence uncertainties, they can also interact.

By interacting locally, the dimensions of communication are integrated (e.g., in the case of an innovation), but this provisional stabilization may contribute to further differentiation at the global level in a next round. Integration and differentiation are two sides of the same coin. The empirical question is which one of the two will prevail. The system of communications, which both differentiates in terms of functions and integrates in terms of performative actions, can be made the subject of a multitude of analytical perspectives (Leydesdorff, 1997, 2001).

**Conclusion**

The crucial question is whether these hypotheses of the self-organization of communication and functional differentiation among the codes of communication add to our understanding of the sciences and the knowledge-based transformations of other social systems (e.g., the economy). I have argued that cognition at the level of the social system is manifested as discursive knowledge which is guided by a latent code of communication. This code of communication remains socially constructed, but it tends to develop into a control mechanism which groups authors and texts selectively. While these groupings can be observed, the cognitions guiding the observables can only be hypothesized.

The declaration of these expectations as potentially different from observations transforms the sociological discourse recursively into one which is guided cognitively. Expectations based on theorizing can be informed by observations. Not incidentally, statistics like chi-square, which test observational values against expectations, have been paradigmatic for the empirical traditions which aim at *explaining* the social beyond its positive manifestations. These statistics have enabled social scientists to test the value of variables for their significance; the algorithms of non-linear dynamics analogously enable us to study *changes* in the values of variables (Δx/Δt; Δy/Δt) in terms of their contribution to the reduction of uncertainty at the systems level.

For example, an endogenous reduction of uncertainty can be considered as a consequence of the self-organizing operation of a system's knowledge base. One can raise the question of whether a specific configuration provides a synergy, for example, in university-industry-government relations. In such a design, however, measurement requires the specification of the cognitive dimension as analytically distinct from the economic or the organizational ones (as in Figure 1 above). The interaction effects can then be positive, neutral or negative, but this remains an empirical question (e.g., Leydesdorff *et al*., 2006). However, one needs additionally a theory of measurement before one can proceed to empirical research.